\documentclass[a4paper,american,floatfix,pdftex,superscriptaddress,twoside,%
aps,pra,
reprint,
]{revtex4-2}%
\usepackage{amsfonts,amsmath,amssymb}
\usepackage[T1]{fontenc}
\usepackage{graphicx}%
\usepackage[utf8]{inputenc}
\usepackage{mathtools}
\usepackage{hyperref, hypernat}
\usepackage[displaymath,textmath,graphics]{preview}
\usepackage{beitmacros}
\usepackage{babel}
\graphicspath{{./figures/}} 

\newcommand{\dds}{\ensuremath{\text{D}_2\text{S}\;}}%

\newcommand{\beit}{\affiliation{BEIT, Mogilska 43 31-545 Kraków, Poland }}

\newcommand{\ezemail}{\email{emil@beit.tech}}%

\begin{document}
\title{Creating rotational coherences in molecules aligned along the intermediate moment of inertia axis. }

\author{Emil J.\ Zak} \ezemail\beit
\thanks{Former affiliation: Center for Free-Electron Laser Science, Deutsches
	Elektronen-Synchrotron DESY, Notkestraße 85, 22607 Hamburg, Germany}

\date{\today}%
\begin{abstract}\noindent%
We propose and computationally study a method for simultaneously orienting the angular momentum of asymmetric top 
molecules 
along:  1) a laboratory-fixed direction; 2)  the molecular intermediate moment of inertia axis; 3) the laser field wavevector. 
For this purpose we utilize a coherent control scheme in which a tailored-pulse optical centrifuge populates rotational
states 
with well 
defined projections of the total angular momentum 
onto molecular axes.  Appropriately time-shaped optical centrifuge pulses can leave the rotational wavepacket in peculiar 
rotational coherences 
which lead to a good degree of 3-dimensional transient alignment, with an 
arbitrary molecular axis 
pointing 
along 
the laser pulse propagation direction. 
As an example, we demonstrate how to generate highly resilient rotational quantum states of \dds in which the molecule 
rotates mainly 
about its intermediate inertia axis, such that its electric dipole moment is permanently aligned along the propagation direction of 
the laser 
pulse.
Applications might include  accessing less obscured
information in various photo-electron imaging experiments.

\end{abstract}
\maketitle%

\section{Introduction}
For gaining  direct access into the physics in the molecular frame, it is imperative to maximally confine at least one of the 
molecule's axes along a laboratory-fixed direction. Experiments utilizing  
high-harmonic generation 
(HHG)~\cite{Spector:NatComm:5:3190,Lock:PRL108:133901,Cireasa2015,piewanowski2014,Woerner:Nature466:604,Baykusheva:PRX8:031060,He2020},
laser-induced electron 
diffraction 
(LIED)~\cite{Blaga:Nature483:194,Corkum:NatPhys3:381,Peters2012,Lin2012,PuthumpallyJoseph2017}, direct 
photo-electrons~\cite{Trabattoni:NatComm11:2546,Johansen:JPB49:205601,Kumarappan:PRL100:093006,Yuan2018,Artemyev2015,Mller2018,Mller2020,Lehmann2013,Demekhin2018,Brausse:PRA97:043429,Kastner2019,Bayer:NJP21:033001,Kastner2020}
or X-ray diffraction~\cite{Kierspel:JCP152:084307,Ho:JCP131:131101} have been successfully used to probe the structure and 
ultra-fast 
dynamics of the gas 
phase molecules. 

However, direct application of these techniques to randomly oriented gas phase molecular samples has limited use, due to the 
averaged-response from all possible laboratory-frame orientations. Information  about the structure and dynamics of 
molecular 
systems obtained from randomly oriented samples is therefore obscured and requires special 
algorithms~\cite{Spector:NatComm:5:3190,Lock:PRL108:133901,Arenz2020} 
 to be reliably retrieved. 
For this reason, several  techniques for determining the molecular structure and dynamics  at 
high temporal 
and/or spatial resolution~\cite{Filsinger:PCCP13:2076, Itatani:Nature432:867,
	Meckel:Science320:1478, Holmegaard:NatPhys6:428, Hensley:PRL109:133202, Barty:ARPC64:415,
	Kuepper:PRL112:083002, Yang:Science361:64,Trippel:PRA89:051401R,Ramakrishna:PRA87:023411}   have greatly benefited 
	from pre-aligning the molecular samples in one~\cite{RoscaPruna:PRL87:153902,
	Chatterley:JCP148:221105, 
	piewanowski2014,Kumarappan:PRL100:093006,Kierspel:JPB48:204002,Karamatskos:NatComm10:3364, 
	Fleischer:PRL107:163603,Trabattoni:NatComm11:2546,He2020,Trippel:PRA89:051401R},
two~\cite{Smeenk:PRL112:253001, Smeenk:JPB46:201001, Korobenko:PRL116:183001}, and
three~\cite{Larsen:PRL85:2470, Tanji:PRA72:063401,Hansen:JPB45:015101, Rouzee:PRA77:043412, Nevo:PCCP11:9912,
	Lin:NatComm9:5134,Viftrup:PRL99:143602} spatial directions. Applications of molecular alignment  have also been found in 
	stereodynamical 
control of chemical
reactions~\cite{Larsen:PRL83:1123, Miranda:NatPhys7:502, Liu:ARPC52:139, Shagam:NatChem7:921}.

Molecular alignment of  linear
and symmetric top molecules in 1-D is routinely performed at high efficiency with strong
laser 
pulses~\cite{Leibscher:PRL90:213001,Karamatskos:NatComm10:3364,Cornaggia:JPB49:19LT01,Hamilton:PRA72:043402,Peronne:PRA70:063410,Hamilton:PRA72:043402,Cryan:PRA80:063412,Larsen:JCP111:7774,Ghafur:NatPhys5:289,Holmegaard:PRL102:023001,Hoshina:JCP118:6211}.
 However, the majority of molecules with more than two atoms are 
asymmetric tops, such as water, to mention one. 
For most imaging experiments it is highly desirable to have the molecule aligned, even transiently,  in space free of any 
fields but the probe, to avoid interference of the aligning field with the probe field. The typically used for this purpose impulsive 
alignment~\cite{Nielsen:PCCP13:18971,Broege:PRA78:035401,Galinis:FD171:195,Rouzee:PRA77:043412,Guerin:PRA77:041404}
approach aims at creating a broad rotational population distribution and a locked-phase relation between the 
wavepacket components. 

Field-free alignment through an impulsive or mixed-field excitation~\cite{Kienitz:CPC17:3740,Omiste:PCCP13:18815,Omiste2016} 
proves 
inherently burdened for 
asymmetric top 
molecules~\cite{Stapelfeldt:RMP75:543,Holmegaard:PRA75:051403R,Seideman:PRL83:4971,Spector:NatComm:5:3190,
	Damari:PRL117:103001}.
 A 
reason for this are three 
different moments of inertia 
along the
principal 
inertia axes, which give incommensurable  frequencies between different rotational quantum states; their 
quantum rotational dynamics is chaotic, meaning the well known perfect quantum revivals observed for symmetric tops are 
no 
longer 
achievable~\cite{Stapelfeldt:RMP75:543,Holmegaard:PRA75:051403R,Seideman:PRL83:4971,Spector:NatComm:5:3190,Viftrup:PRL99:143602}.
As there is no obvious mitigation to this problem, experiments~\cite{Spector:NatComm:5:3190,Lock:PRL108:133901} have aimed 
at kicking the 
asymmetric top molecule with an ultra-short intense non-resonant laser pulse and decomposing 
the measured response onto the 
three perpendicular principal inertia axis contributions - usually very low degrees of transient  alignment  of 
smallest, middle or largest moment of inertia axis have been 
observed~\cite{Rouzee:PRA77:043412,Larsen:PRL85:2470,Artamonov:MP110:885,Joireman:JCP96:4118,Felker:JPC90:724,Connell:JCP94:4668}.
 
 \begin{figure}
 	\includegraphics[width=1.0\linewidth]{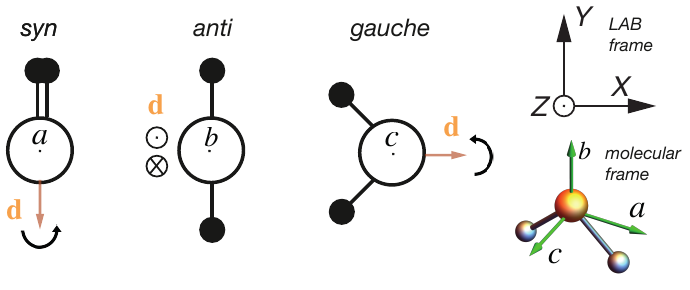}%
 	\caption{Schematic view of an asymmetric top molecule rotating about its $a$-, $b$- and $c$- principal inertia axis. From the 
 		laboratory frame perspective the molecules are in \textit{syn}, \textit{anti} and \textit{gauche} geometry. We assume that the 
 		probe pulse propagates along the laboratory $Z$-direction, whereas the molecule-confining pulse is restricted to the $XY$ 
 		plane. $d$ denotes the dipole moment vector. }
 	\label{fig:scheme}
 \end{figure}
 
Mastering the alignment of the smallest ($a$-), middle ($b$-) and largest ($c$-) polarisability axis along a laboratory-fixed 
 direction would open a 
 vast field of possibilities for imaging the molecular-frame dynamics, with direct tomography of molecular 
 orbitals~\cite{Itatani:Nature432:867,Spector:NatComm:5:3190,Ramakrishna:PRA87:023411,Maurer:PRL109:123001,Graus2019} as 
 an example. 

Three major types of molecular axis alignment (angular momentum orientation)  are shown in~\autoref{fig:scheme}.
On the example of a bent triatomic 
molecule,~\autoref{fig:scheme} displays possible orientations of the molecular polarisability axis with respect to 
the laboratory $Z$-direction:\textit{ syn} ($a$-axis aligned with $Z$), \textit{anti} ($b$-axis aligned with $Z$) and 
\textit{gauch}e 
($c$-axis aligned with $Z$).  
In the \textit{syn} and \textit{gauche} case, the intermediate polarisability axis (to a good approximation the intermediate inertia 
axis, 
very often also the dipole moment) of the molecule 
rotates in the plane of the confining field, whereas  in the \textit{anti} orientation, the dipole moment permanently points along the 
wavevector of the confining field ($Z$-direction).
A missing piece to complete the picture of molecular alignment geometries is placing the symmetry axis, or more generally 
the intermediate inertia $b$-axis of an asymmetric top 
molecule, along the probe pulse's wave-vector; i.e. achieving the  \textit{anti}  alignment shown in~\autoref{fig:scheme}. 
Such an alignment is complementary to the $a$- and $c$-axis alignments.

Whereas the entire class of impulsive and mixed-field alignment techniques rely on strong confinement of the spatial 
observables, at the cost of high uncertainty in the angular momentum, an alternative approach of the angular momentum 
orientation~\cite{Zhdanovich:PRL107:243004, 
	Milner:ACP159:395, Milner:PRA93:053408, Owens:PRL121:193201} utilizes a reversed principle. It gives up spatial 
confinement in two directions ($x$,$y$), to strongly orient the angular momentum z-component $J_z$ and the $z$-axis 
along the same laboratory-frame direction.
An advantage carried by large magnitude and strongly oriented angular 
momentum samples is their stability and high resistance to collisional decoherence for up to
microseconds~\cite{Yuan:PNAS108:6872, Khodorkovsky:NatComm6:7791, Milner:PRL113:043005,
	Milner:PRX5:031041}. Such resilient molecular gyroscopes can have either  their largest-  or  smallest- electronic 
polarisability axis aligned with the polarisation plane of the confining electric 
field~\cite{zak2021,Owens:JPCL9:4206,Korobenko:PRL116:183001}. 

So far,  alignment of the intermediate ($b$-) 
polarisability axis have remained elusive both theoretically and experimentally. None of available impulsive or adiabatic alignment 
or
angular momentum 
orientation techniques were proposed to realize the \textit{anti} geometry from~\autoref{fig:scheme}.
This is because the rotation about the intermediate axis (\textit{b-rotation}) is quite different from the other two rotations, both 
classically 
and 
quantum-mechanically. 

Classically the $b$-rotation is unstable, with the molecule (object) flipping its orientation 
periodically~\cite{Ashbaugh1991,Zhilinskii1999,Hamraoui2018,Damme2017,Ma2020}, which is also known as the 
tennis racket theorem or the Dzhabenikov effect~\cite{Ma2020,Hamraoui2018}: the intermediate axis lies on the 
phase-space separatrix between two stable points of $a$- and $c$- rotation. 

\begin{figure*}
	\includegraphics[width=1.0\linewidth]{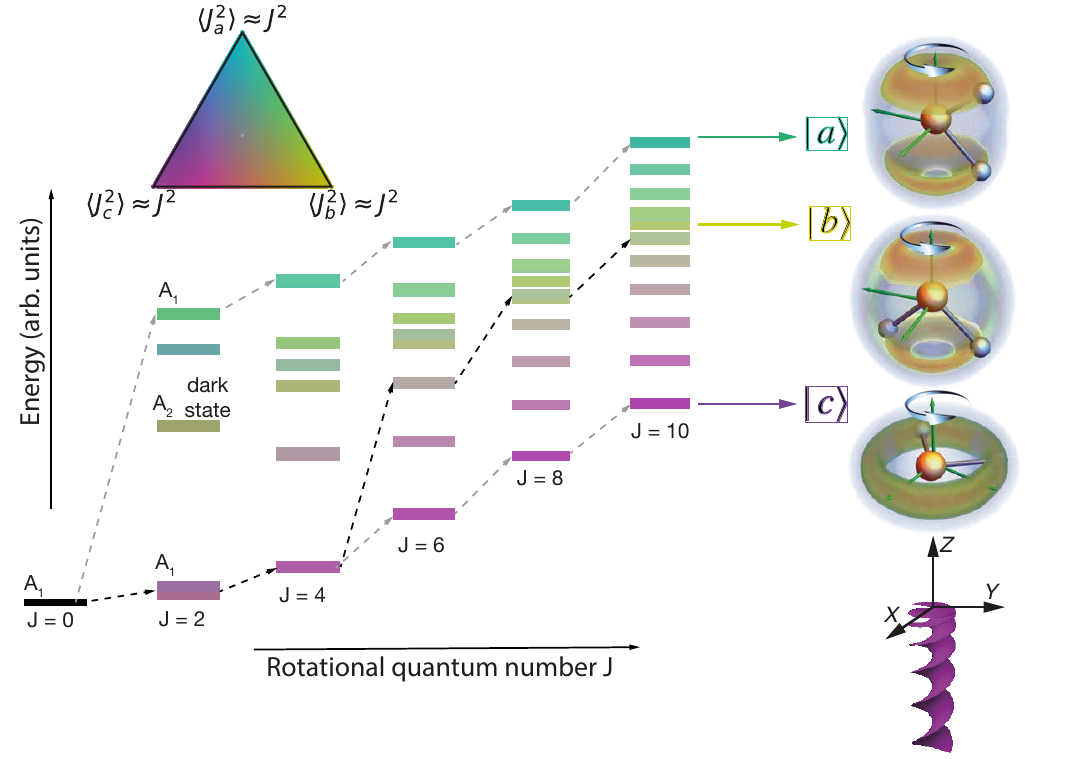}%
	\caption{Rotational energy levels of \dds and a selected rotational excitation path leading to the principal rotation $b$-states 
		$|b\rangle$ marked with dashed black arrows. For
		any value of the rotational quantum number $J$ there is a multiplet of $2J+1$ levels, which are
		colored according to the average values of the angular momentum projection operators onto the \dds's
		principal axes of inertia $\expectation{\hat{J}_a^2}$ (cyan color), $\expectation{\hat{J}_b^2}$ (yellow color), and
		$\expectation{\hat{J}_c^2}$ (purple color). A continuous map covering all possible angular momentum projections is shown in 
		the 
		triangle in the inset. Right panel shows 3D probability densities for finding deuterium nuclei in three different principal rotation 
		states ($|a\rangle$, $|b\rangle$ and $|c\rangle$. The laboratory frame and a schematic of an optical centrifuge pulse are 
		shown in the bottom right corner.)
	}
	\label{fig:energies}
\end{figure*}

Quantum mechanically, the situation is quite 
different. First, as long as no parity-breaking interaction is present, the molecule may only be aligned along with the external field, 
such that no 
preferable orientation is distinguished. In this sense, the 
flipping-rotation known from classical physics is inherently present in the quantum wavefunction. Secondly, 
rotational states attributed to the rotation about the $b$-axis are located in the middle of the rotational energy levels 
manifold, making it cumbersome to access via optical methods (see~\autoref{fig:energies}), in contrast to accessing the $a$- or 
$c$- 
rotating 
states.
So far, there have been no reports of a coherent optical excitation targeting these $b$-rotation states. 

In this work, we show how to control the rotation axis of asymmetric top molecules simultaneously in the laboratory-frame and in 
the 
molecular-frame, with the classically unstable  axis of rotation aligned with the laser field wave-vector.

First we computationally identify and characterize rotational states of asymmetric top molecules, which can be 
associated with the rotation  about the intermediate inertia axis ($b$-rotation states); they lie in the middle of the rotational 
energy levels 
manifold. Next, we utilize a modification of the optical centrifuge technique already introduced in Ref.~\cite{zak:dyn-chirality} to 
show 
how the $b$-rotation states can be efficiently populated in the asymmetric top \dds molecule. Appropriate timing of the centrifuge 
pulse creates arbitrary coherences between the stationary $b$-axis rotating states, some of which exhibit classical-like rotational 
motion~\cite{Lapert:PRA83:013403,Korobenko:PCCP17:951} and a fair degree of 3D alignment, in field-free conditions. A new 
type of rotational  transient, \textit{B-type}, is discussed. 

The results presented in this work, together with the results of Ref.~\cite{zak2021} complete the study of 1D- and 
3D-alignment of asymmetric top molecules (through angular momentum orientation), with either of their $a$-, $b$- or $c$- 
polarisability axis pointing along the 
wavevector of the probing pulse~\cite{Smeenk:JPB46:201001,Koch:RMP91:035005}. In principle, complete molecular-frame 
information could be collected in an imaging experiment combined with the technique described below. 

\section{Principal rotation states of \dds}
\autoref{fig:energies} displays a schematic of rotational energy levels of the \dds molecule with low total angular momentum . For
any value of the rotational quantum number $J$ there is a multiplet of $2J+1$ levels, which are
colored according to the average values of the angular momentum projection operators onto the \dds's
principal axes of inertia $\expectation{\hat{J}_a^2}$ (cyan color), $\expectation{\hat{J}_b^2}$ (yellow color), and
$\expectation{\hat{J}_c^2}$ (purple color). A continuous map covering all possible angular momentum projections is shown in the 
triangle in the inset to~\autoref{fig:energies}.

Highest-energy levels at each total angular momentum $J$
correspond to $\expectation{\hat{J}_a^2}\approx J^2$ (cyan color), \ie, in these states
the molecule rotates about the $a$-axis and $K_a=J$ becomes a near-good quantum number, denoting the projection of the total 
angular momentum onto the $a$ principal-inertia-axis. Such state can be dubbed a \textit{principal rotation state} and marked 
with 
$|a\rangle$. 
Similarly, the
lowest-energy levels correspond to rotation about the $c$-axis with $K_c=J$ (purple color), while those with energies in the 
middle are 
mixtures of
rotations about different axes with some of them exhibiting classically unstable $b$-axis rotation
(yellow color).

In the right panel in~\autoref{fig:energies} shown are the calculated 3D
probability densities for the deuterium atoms for the highest-, middle-, and lowest-energy levels at
$J=10, M=10$. Here, $M$ is the quantum number for the $Z$-component of the angular momentum operator
in the laboratory-fixed frame, which is shown in the bottom right corner, together with an optical centrifuge pulse.
A non-resonant rotational excitation leading to the population of  principal rotation states $|a\rangle$, 
$|b\rangle$ or $|c\rangle$ can be realized by the interaction with linearly polarized light with accelerated rotation of its 
polarization plane, 
known as an optical centrifuge~\cite{Korobenko:PRL116:183001,zak2021,Owens:JPCL9:4206}. Interaction of the electronic 
polarisability with the centrifuge 
pulse promotes only transitions with $\Delta{J}=2$ and $\Delta{M}=\pm2$, depending on the handiness of the 
centrifuge pulse.
Lowest rotational energy  excitation path  ($c$-path marked with thin gray arrows in~\autoref{fig:energies} ) proceeds in 
near-oblate 
asymmetric top molecules, such as \dds, at low acceleration rates of the optical 
centrifuge pulse, as discussed in Ref.~\cite{zak2021}. High acceleration rates of the centrifuge field populate 
predominantly states along the highest excitation path ($a$-path). In near-prolate molecules, populating states along the 
$c$-path requires appropriate modulation of the centrifuge's field intensity~\cite{Owens:JPCL9:4206}.

\section{Populating b-principal rotation states with an optical centrifuge}
A proposition to coherently populate $|b\rangle$ principal rotation states has been so far lacking. Here we propose how to use the 
optical centrifuge to populate the $|b\rangle$ states.
Because such an excitation cannot come directly through the $|J=0,M=0,+ \rangle \rightarrow |J=2_{K_b \approx 
2},M=2,+\rangle$ 
transition, as the 
upper state has $A_2$ symmetry, which in the absence of parity-breaking (electric dipole) interaction remains dark, 
another excitation path must be found. 
For this purpose we utilized a  shortest path finding algorithm~\cite{zak:dyn-chirality} to determine an optimal excitation pathway 
(based on largest overall
transition moment) connecting the ground state with a $|b\rangle$-state at $J=14, M=14$. Initial part of this path is shown 
in~\autoref{fig:energies} with thick black dashed arrows and it reads $|J,M,h,\tau \rangle$:
\begin{equation}\small
\begin{split}
&|0,0,1,+ \rangle \rightarrow  |2,2,1,+ \rangle \rightarrow  |4,4,1,+ \rangle \rightarrow 
	 |6,6,5,+ \rangle \rightarrow  \\&|8,8,9,+ \rangle  \rightarrow  |10,10,13,+ \rangle \rightarrow  |12,12,17,+ \rangle  \rightarrow  
	 |14,14,21,+ \rangle 
\end{split}
\label{eq:path}
\end{equation}
where $h$ labels the rotational states in ascending energy order for each $J$, whereas $\tau$ is the rotational parity.
With the rotational excitation path selected, the optical centrifuge pulse envelope has to be modulated to reach significant 
intensities
near 
the times at which the pulse's instantaneous frequency matches resonance with the rotational transition. 
The intensity envelope of the centrifuge pulse
has been modeled with \textit{sinc} functions centered at appropriate resonance times determined by the centrifuge's acceleration 
rate $\beta$. Such tailoring of the centrifuge's pulse 
intensity is 
quite 
universal, so that both near-prolate molecules and near-oblate molecules 
can be efficiently excited into their $b$-rotation states. Intensity-modulated pulse profiles can now be created with the use 
of a 4-f pulse shaper combined with a standard centrifuge 
setup~\cite{Kerbstadt:shaping:inprep}. 

The time-dependent Schr{\"o}dinger equation has been solved on a grid of the
centrifuge parameters: field strength $E_0$, width of sinc function $\sigma$ and acceleration rate $\beta$ to find optimal and 
robust values: $E_0 = 1.7 \cdot 10^8$~V/cm,  $\sigma= 7$~ps and  $\beta= 60$~GHz/ps, which guarantee good excitation 
efficiency. 
The rotational-dynamics calculations of \dds were performed in two steps. First, energies and their transition moments were 
obtained with the full-dimensional 
variational 
procedure TROVE~\cite{Yurchenko:JMS245:126,
	Yachmenev:JCP143:014105, Yurchenko:JCTC13:4368} together with a highly-accurate spectroscopically
adjusted potential energy surface~\cite{Azzam:MNRAS460:4063} and a high-level \emph{ab initio}
polarizability surface~\cite{Owens:JPCL9:4206} of the H$_2$S molecule within the Born-Oppenheimer
approximation.  In the second step, the time-dependent solutions
for the full molecule-field interaction Hamiltonian were obtained using the computational approach
Richmol~\cite{Owens:JCP148:124102,Yachmenev:JCP151:244118} in the field-free basis.  The wavefunctions were 
time-propagated
using the split-operator method with a timestep of 10~fs. The time-evolution operator was evaluated
using an iterative procedure based on the Krylov subspace methods.

\begin{figure}
	\includegraphics[width=1.0\linewidth]{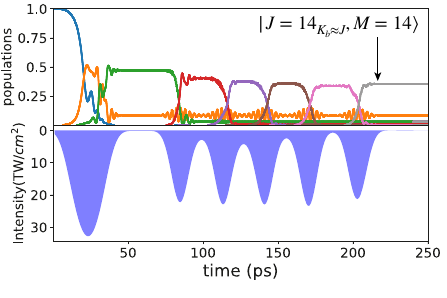}%
	\caption{Time-profiles of \dds rotational states' populations along the excitation path given 
	in~\autoref{eq:path} leading to a $b$-principal rotation state. Bottom panel:  
	intensity profile of the  
		centrifuge 
		pulse.  Target state $|J=14_{K_b \approx J},M=14\rangle$ is marked with an arrow.}
	\label{fig:populations}
\end{figure}

\autoref{fig:populations} displays time-profiles of rotational state populations of \dds along the selected \textit{b-rotation} 
excitation path given in~\autoref{eq:path}. Starting from the rotational ground state (see~\autoref{fig:energies}), the excitation 
proceeds at a good 
yield 
($\approx 35\%$) up 
to 
the target state $|J=14_{K_b \approx J},M=14\rangle$. Bottom panel in~\autoref{fig:populations} shows the intensity profile of 
the  
centrifuge 
pulse. The major source of population losses occurs at the initial forking stage  $J=0,M=0 \rightarrow 
J=2,M=2$. The efficiency can be enhanced by lowering the centrifuge's acceleration rate, at the cost of a longer 
pulse. 

\section{Properties of b-principal rotation states}
\begin{figure}
	\includegraphics[width=1.0\linewidth]{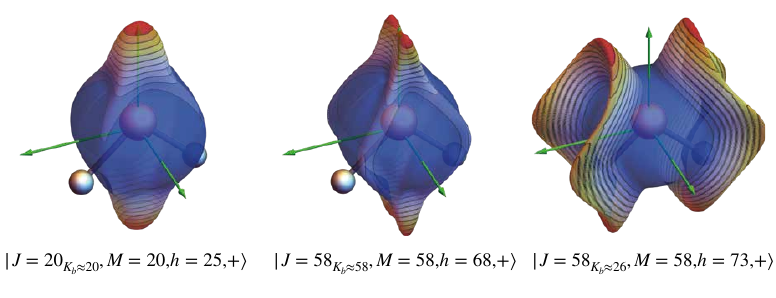}%
	\caption{3D	probability distributions for the molecular-frame rotation axes of \dds, plotted for three different 
		$|b\rangle$-principal rotation states. $|J_{k_b},M,h,\tau\rangle$ denotes the rotational state with $h$ labelling 
		energy levels in ascending 
		order for each $J$-multiplet, and $\tau$ is parity. $M=J$ implicitly. Green axis system denotes the molecular principal inertia 
		frame.}
	\label{fig:axis-densities}
\end{figure}

\autoref{fig:axis-densities} depicts 3D	probability distributions for the molecular-frame rotation 
axes of \dds, plotted for three 
different 
$|b\rangle$-principal rotation states. The rotational states are denoted as $|J_{k_b\approx J},M,h,\tau\rangle$, with $h$ labelling 
energy levels in ascending 
order for each $J$-multiplet, and $\tau$ is parity. 
We selected a representative $|b\rangle$-principal rotation state  $|20_{k_b\approx 20},20,25,+\rangle$
shown in the left panel in~\autoref{fig:axis-densities} to demonstrate how the rotation-axis probability density
is localized near the $b$-axis. 

 The degree of 1D alignment of the rotation axis along the $b$-axis 
is similar for all $|b\rangle$ principal rotation states and reads $\langle \cos^2 
\theta_{bZX} \rangle \approx 0.93$, with little sensitivity to $J$. The $|b\rangle$ states can be found up to high values of the total 
angular 
momentum $J$,
with $|58_{k_b\approx 58},58,68,+\rangle$ state shown in the middle panel  in~\autoref{fig:axis-densities}. 
This state exhibits a sharp maximum probability which  splits into 
two poles located on opposite sides of the $b$-axis, at $\theta =9.1^o, 
\phi=90.5^o,269.7^o$ and $\theta =170.1^o, 
\phi=90.5^o,269.7^o$. The degree of alignment for the $|58_{k_b\approx 58},58,68,+\rangle$  state measured along the $b$-axis 
reads $\cos^2 
\theta_{bZX} = 0.91$.
The sharpening of the 3D rotation axis probability maximum with increasing $J$ does not necessarily translate into higher 
degrees of 
alignment, due to emergence of new probability islands related to rotational energy level clustering. This 
clustering effect causing rearrangement of rotation axis has been discussed 
elsewhere~\cite{zak:dyn-chirality,Owens:JPCL9:4206}. 

Interestingly some cluster states located near the middle of a given  $J$-rotational energy manifold have their
probability density for the rotation axis lying close to the classical 
separatrix~\cite{Ashbaugh1991,Zhilinskii1999,Hamraoui2018,Damme2017,Ma2020}
between the $a$- and $c$- stable rotation, as the one shown in the right panel in~\autoref{fig:axis-densities}. The probability for 
the axis of rotation is then delocalized along a well defined path connecting opposite orientations of the molecule.
Clasically, rotation of the molecule about $a$- or $c$- axis or near these axes features stable \textit{oscillating} or \textit{rotating}
trajectories~\cite{Damme2017} of the molecule-fixed angular momentum. Rotation axis initially placed near the $b$-axis is 
unstable, showing periodic flipping, whose period is sensitive to the initial conditions. This latter behavior is 
known as the \textit{tennis racket effect}~\cite{Ashbaugh1991,Zhilinskii1999,Hamraoui2018,Damme2017,Ma2020}.
Quantum mechanically, the stationary $|a\rangle$, $|b\rangle$ or  $|c\rangle$ states have an unchanged probability
density as well as both orientations are equivalent, such that axis flipping is inherent. Orientation of such states
is difficult via the interaction with the electric dc field due to large energy separation 
between opposite parity pairs ($A_1/A_2$). 

One can create transient orientation of principal rotation states through coherent superposition of opposite parity states, for 
instance: $|A_1\rangle + e^{-i\Delta E 
/\hbar t} |A_2\rangle$, in which 
the frequency of the dipole moment vector flipping equals the energy $\Delta E$ separating  the $|A_1\rangle , |A_2\rangle $ 
states. If this 
separation is large, the oscillations are very fast. We observe that all types of principal rotation states, that is $|a\rangle$, 
$|b\rangle$ and
$|c\rangle$ have their respective $|A_1\rangle$ counterparts located within several wavenumbers, such 
that a unidirectionally rotating and oriented states can in principle be created. However in  the field-free case, all such 
wavepackets will exhibit 
flipping of the 
dipole moment at a rate given by the energy splitting between the  $|A_1\rangle , |A_2\rangle $  states. 
In this sense, the tennis racket effect is present in all quantum  rotational states. 
Another possibility is to use a  strong dc field  to create orientation, in which case it is possible to create $|A_1\rangle \pm 
|A_2\rangle$ field-dressed states with locked 
phase 
between its 
components, such that the molecule is oriented up or down, respectively. Such orientation lasts as long as the field is present. 
Very strong dc electric fields up to few MV/cm are needed to achieve orientation this way. 
 
 \section{Coherences between principal rotation states: road to 3D alignment}
\begin{figure*}
	\includegraphics[width=1.0\linewidth]{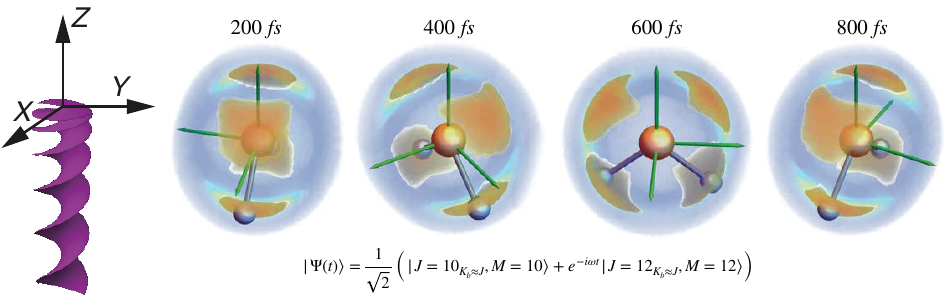}%
	\caption{Snapshots of 3D probability density of the deuterium atoms in a \textit{cogwheel state} of \dds: 
	$\frac{1}{\sqrt{2}}\left( 
	|10, k_{b}\approx 10, 10,+\rangle + e^{-i\omega t} |12, k_{b}\approx 12, 12,+\rangle  \right)$ . Transient 3D alignment is 
		visible.  In the left panel the $XYZ$ laboratory-frame is marked together with the centrifuge pulse. The period of rotation for 
		this state is $T=154$fs.}
	\label{fig:cogwheel}
\end{figure*}

As shown in~\autoref{fig:populations}, appropriate turn-off time of the centrifuge pulse can leave the wavepacket in a  
coherence 
between neighboring $J$ and $J+2$ rotational states. An interesting characteristic of such two-state wavepackets is a 
classical-like 
(cogwheel) rotation of the rotational probability density~\cite{Lapert:PRA83:013403,Korobenko:PCCP17:951}. Interference 
between the $|b_J\rangle$ and $|b_{J+2}\rangle$ state 
spawns a well localized nuclear density, which revolves about the $b$-axis at a constant frequency determined by the energy 
difference 
between the composing states. \autoref{fig:cogwheel} displays four snapshots of such a probability density for the deuterium 
atoms 
in \dds for the 	$\frac{1}{\sqrt{2}}\left( 
|10, k_{b}\approx 10, 10,+\rangle + e^{-i\omega t} |12, k_{b}\approx 12, 12,+\rangle  \right)$ coherence. The classical-like rotation 
period 
decreases 
linearly with the total angular momentum $J$ and for the $J=10,12$ pair it is $T=154$~fs which corresponds to 216~\invcm 
energy separation 
between states composing the cogwheel wavepacket. 

Effectively, the cogwheel wavepackets feature highly oriented angular momentum along the molecule-fixed $z$-axis and the 
laboratory-fixed $Z$ axis as well as a degree of molecular-axis localization in the plane perpendicular to the angular momentum: 
the $xy$ and $XY$ plane.
The degree of 3D alignment in a cogwheel principal rotation 
state can be measured with the 2D alignment cosines of respective molecular axes onto the laboratory $XZ$ and $XY$ planes, 
which in the case of wavepacket from~\autoref{fig:cogwheel} are given by: $\langle\cos^2 
\theta_{bZX}\rangle = 0.71-0.81$, $\langle\cos^2 \theta_{cZX}\rangle = 0.18 - 0.4$, $\langle \cos^2 
\theta_{cXY}\rangle = 0.28 - 0.73$. 

Both the probability densities 
shown in~\autoref{fig:cogwheel}  
and the 
alignment cosines were calculated using monte-carlo 
sampling (1 Million points) of the rotational wavefunction. We must note here that \dds, which has been chosen for computational 
convenience, is a molecule with very large rotational 
constants ($A = 164.57$~GHz, $B = 135.38$~GHz, $C = 73.24$~GHz)~\cite{NIST:CCCBDB}, meaning that the rotational 
dynamics is
also relatively fast. Heavier rotors, such as SO$_2$ or benzene derivatives  have their cogwheel-rotation-period
in the range of several picoseconds. 

The probability density for finding the deuterium atoms in 3D space in a cogwheel $J$,$J$+2 state, such as the one depicted 
in~\autoref{fig:cogwheel}, can be written as:
\begin{equation}
\begin{split}
\rho_J(\theta,\phi,\chi,t) &= N \cos ^{4J}\frac{\tilde{\theta}}{2}\left[2J+1 + (2J+5)\cos ^{8}\frac{\tilde{\theta}}{2}\right. \\
&\left.+2\sqrt{2J+5}\cos\left( \Delta J(\tilde{\phi}+\tilde{\chi})-\omega t\right)\right]
\end{split}
\label{eq:cog}
\end{equation}
 where $\tilde{\theta} = \theta - 
\theta_D$,  
$\tilde{\phi} = \phi- \phi_D$ and  $\tilde{\chi} = \chi- \chi_D$ are the Euler angles denoting the position of deuterium atoms in \dds, 
in bisector molecular-frame embedding, i.e. where 
the 
$b$-axis lies along the molecule-fixed $z$-axis. $N$ is the normalization constant.

The functional 
form of the cogwheel probability density shown in~\eqref{eq:cog} suggests that 
cogwheel states have at most two 
effective degrees of rotational freedom: $\theta, \phi+\chi$ and that the nuclear density rotates about the molecule-fixed $z$- 
and 
laboratory-fixed $Z$-axis at a frequency $\omega$. In~eq.\eqref{eq:cog} $\Delta J = 2$, such that the cogwheel has two 'teeth', 
which 
also fixes the maximum degree of transient 3D alignment to $\langle \cos^2 \theta_{aXY}\rangle = \langle \cos^2 
\theta_{cXY}\rangle =  0.73, \langle \cos^2 
\theta_{bXZ}\rangle = 0.93 $.

A higher degrees of field-free alignment could be achieved if one created a coherent wavepacket in which one molecular axis 
is strongly confined to the laboratory $Z$-axis, while the wavefunction projected onto the plane perpendicular to $Z$ produced 
constructive interference at given rotational revival times.
Such a scenario is possible with a coherence between several principal rotation states $a$-, $b$- or $c$- , as 
written in the equation 
below:
\begin{equation}
\begin{split}
\psi(t) =&  \sum_{J=J_{min}}^{J_{max}} c_J |J_{K_{\lambda}\approx J}, M=J,h_J,\tau\rangle \approx\\
& \approx \sum_{J=J_{min}}^{J_{max}} |c_J|e^{-i(\Delta 
J(\phi+\chi)-E_Jt+\phi^0_J)}\cos^{2 J}\frac{\theta}{2} +h.c.
\end{split}
\label{eq:cogwf}
\end{equation}
where $\lambda = a,b$ or $c$. Here $\phi^0_J$ is the static phase of the $c_J$ complex coefficient and $\Delta J=2$ denotes the 
difference in the total angular momentum between states in the wavepacket. 

The idea behind the wavepakcet shown in~\eqref{eq:cogwf} is to create rotational 
transients~\cite{Joireman:JCP96:4118,Felker:JPC90:724,Connell:JCP94:4668,Artamonov:MP110:885,Artamonov2008,
	Viftrup:PRA79:023404,Tenney2016},
 in the 
coherent superposition of the  principal rotation states. The alignment cosine measuring the degree of alignment in the laboratory 
$XY$ plane with respect to the $X$-axis  can be calculated analytically for the coherence given in~\eqref{eq:cogwf}:
\begin{equation}
\begin{split}
\langle \cos^2\phi \rangle _{\psi(t)} = \sum_{J =J_{min}}^{J_{max}-2} g_{J} \cos\left(\omega_{J+2J}t + 
\Delta\phi_{J+2J}^0\right)
\end{split}
\label{eq:cogcosine}
\end{equation}
where $g_{J}  = \frac{1}{4J+6} |c_J| |c_{J+2}| \sqrt{(2J+1)(2J+5)} $. In derivation of~\eqref{eq:cogcosine} we assumed that 
$K_\lambda = J$, 
and that $\lambda = a, b$ or $c$ is the molecule-fixed $z$-axis. Full expression for the 
alignment cosine, in a non-perfect principal rotation state is given in~\autoref{sec:appendixA}. 

From the form of the wavefunction given in~\eqref{eq:cogwf} we infer that the degree of alignment of the molecular axis 
$\lambda = a, b$ or $c$ along the laboratory-fixed $Z$-axis is high, depending which principal rotation state one pick: 
$|a\rangle$, $|b\rangle$ or  $|c\rangle$, respectively. 
The other two molecular axes are confined to the $XY$ plane and when the coherence given in~\eqref{eq:cogwf}  consists of 
more than two rotational states, the $\langle \cos^2\phi \rangle _{\psi(t)} $ alignment cosine exhibits a revival pattern. 
This suggests that at certain times the degree of alignment with respect to the laboratory $X$-axis bound to reach maximum, 
leading to 
transient 3D alignment. 
\begin{figure}
	\includegraphics[width=1.0\linewidth]{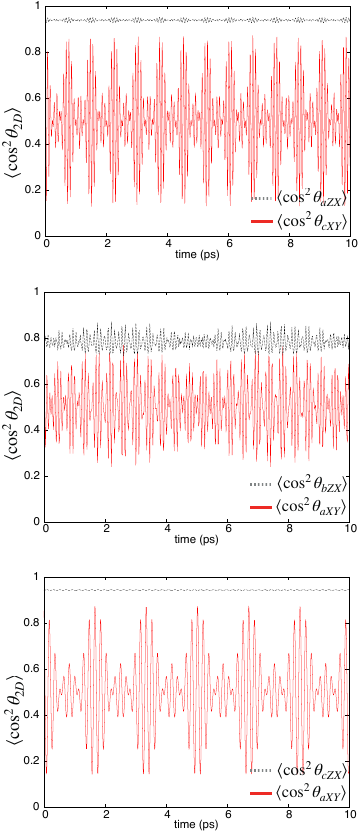}%
	\caption{Calculated 2D alignment-cosine time-profiles for three wavepackets, representing coherences between the 
		$a$-(uppermost panel), $b$-(middle panel)
		and $c$-(bottom panel) principal rotation states in \dds in the range $J=14-20$ (see~\eqref{eq:cogwf}).  The alignment 
		cosines measure
		the degree of alignment of the principal axes of inertia of \dds with respect to the
		laboratory axes. For instance $\langle \cos^2 \theta_{aZX}\rangle $ is the 2D alignment
		cosine calculated with respect to the laboratory $Z$ axis of the projection of the molecular
		$a$ axis onto the $ZX$ detector plane. High degree of permanent 1D alignment is visible as well as sharp revivals of 3D 
		alignment are present in all three cases. }
	\label{fig:cosines}
\end{figure}
To demonstrate this concept, we calculated the alignment cosines for the principal inertia axes of \dds in the rotational state 
given 
in~\eqref{eq:cogwf} with 
$J_{min} 
= 14$ and $J_{max} = 20$ for $\lambda = a$ and a high degree of 3D alignment was reached:
$\langle \cos^2 \theta_{cXY}\rangle_{max} = \langle \cos^2 \theta_{bXY} \rangle_{max} = \langle \cos^2\phi \rangle_{max} = 0.88, 
\langle \cos^2 
\theta_{aXZ} \rangle_{max} = 0.94 $, due to an A-transient~\cite{Felker:JPC90:724,Tenney2016}. A flat relative-initial phase 
relation between the respective components 
was assumed. 
Interestingly, 
a choice of random phases also yields high degrees of maximum alignment, with $\langle\cos^2 \theta_{cXY} \rangle_{max} 
\approx 0.85$. 

Generation of a C-type and most interestingly a \textit{B-type} transient is depicted in~\autoref{fig:cosines}, where time-profiles 
were calculated for appropriate 2D alignment cosines for three wavepackets, representing coherences between the 
$a$-(uppermost panel), $b$-(middle panel)
and $c$-(bottom panel) principal rotation states in \dds in the range $J=14-20$ (see~\eqref{eq:cogwf}). 

The top panel shows how the $\langle\cos^2 \theta_{aXZ}\rangle(t)$ alignment cosine maintains a high value of about 0.94 while 
the $\langle \cos^2 
\theta_{cXY}\rangle$ 
alignment cosine has periodic maxima, reaching up to 0.88.  Similarly the bottom panel in~\autoref{fig:cosines} shows how the 
$c$-principal inertia axis is confined to the laboratory $Z$ axis ($\langle \cos^2 \theta_{cZX}\rangle_{max}$=0.94) and the 
$a$-axis 
shows signs of transient alignment along the $X$ axis ($\langle \cos^2 \theta_{aXY}\rangle_{max}$=0.87).

In the middle panel in~\autoref{fig:cosines} we can see a new type of transient~\cite{Felker:JPC90:724,Tenney2016}:   the B-type, 
which corresponds to $\Delta J = 2$ 
and $\Delta K = 1$ coupling of rotational states ($|b\rangle \leftrightarrow |b'\rangle $). Coherences between the principal rotation 
$|b\rangle$ states reveal a fair 
transient degree of 3D alignment with $\langle \cos^2 \theta_{bZX}\rangle_{max}\approx 0.8$ and $\langle \cos^2 
\theta_{aXY}\rangle_{max}\approx 0.78$. For consistency, we have not chosen for the middle panel in ~\autoref{fig:cosines}  a 
coherence between $b$-principal rotation 
states which 
produced highest possible degree of alignment (1D or 3D), but rather kept identical $J$-range ($J=14-20$) across all panels.

A 3D alignment of molecular axes and 1D orientation of angular momentum, as shown in~\autoref{fig:cosines},  can be generated 
with appropriately tailored  
optical centrifuge pulses, with the pulse 
area and timing calculated with the aid of coherent optimal control~\cite{Ortigoso1998,Salomon2005,Coudert2018}.

To summarize, we demonstrated the existence of rotational quantum states in asymmetric top molecules in which the rotation axis 
is nearly aligned with the intermediate inertia axis. We call them principal rotation states. In a computational study, an optical 
centrifuge 
technique was shown to efficiently populate a selected $|b\rangle$ principal rotation state, in which the angular momentum is 
fixed both along the laboratory-fixed $Z$ axis and the molecular $b$-axis and along the laser field's wavevector. Such an 
alignment is complementary to the previously 
considered alignments of the most- and least- polarisable molecular axis. Interestingly, in \dds the \textit{b-alignment} results in 
the molecule's dipole moment permanently aligned with the wavevector of the confining field and of the following probe pulse, in 
some analogy to \textit{k-alignment} introduced by Smeenk et. al.~\cite{Smeenk:PRL112:253001, Smeenk:JPB46:201001}

Finally, we also presented a scheme allowing to achieve a good degree of transient 3D alignment, with an arbitrary molecular axis 
pointing 
along 
the laser
field's wavevector. The accompanying strongly oriented angular momentum of nearly arbitrary magnitude preserves the created 
coherence, in the same way as gyroscopes do, giving an elevated resilience to collisional decoherence, hence adding stability to 
the degree of alignment. 
Applications of the 3D $a-,b-$ and $c-$alignment can enable access to less obscured
information in various photo-electron imaging 
experiment~\cite{Itatani:Nature432:867,Spector:NatComm:5:3190,Ramakrishna:PRA87:023411,Graus2019},

\section*{Acknowledgments}
I thank Jochen Küpper and Andrey Yachmenev  for helpful discussions. 

\section*{Disclaimer}
Part of this manuscript has been written during the time of  the author's stay at  the Center for Free-Electron Laser Science, Deutsches
Elektronen-Synchrotron DESY, although no part of this work has been created in the time of performing duties included in the author's employment contract. Therefore this work is not an official publication from the Center for Free-Electron Laser Science or the Deutsches Elektronen-Synchrotron DESY.

\section{appendix A}
\label{sec:appendixA}
Below a derivation is given of the equation for the alignment cosine from~\eqref{eq:cogcosine}. We start with a an even parity
asymmetric top rotational state which has possibly maximal projection of the total angular momentum onto the quantization axis 
$\lambda = a$,$b$ or $c$:
\begin{equation}
\begin{split}
\psi(t) = \sum_{J=J_{min}}^{J_{max}} c_J |J_{K_{\lambda}\approx J}, M=J,h_J,\tau\rangle
\end{split}
\label{eq:asymwf}
\end{equation}
where each contributing state is expanded in the symmetric-top basis:
\begin{equation}
\begin{split}
 |J_{K_{\lambda}\approx J}, M=J,h_J,\tau\rangle = \sum_{K=0}^J a^{J,h_J,\tau}_K|J,K,M=J,\tau \rangle
\end{split}
\label{eq:asymwf2}
\end{equation}
and
\begin{equation}
\begin{split}
|J,K,M,\tau \rangle = \frac{1}{\sqrt{2}}\left[|J,k,M\rangle +(-1)^{\tau}|J,-k,M\rangle \right]
\end{split}
\label{eq:symwf}
\end{equation}
with $|J,k,M\rangle=\sqrt{\frac{2J+1}{8\pi^2}}D^{(J)*}_{MK}(\theta,\phi,\chi)$ being the standard symmetric-top wavefunctions. 
Note that the value of 
$K_{\lambda}$ 
in~\eqref{eq:asymwf} depends on the molecular-frame embedding. We choose $\lambda =z$, such that $K_{\lambda} \approx 
J$ for $\lambda = a,b,c$. Then approximately one can write $|J_{K_{\lambda}\approx J}, M=J,h_J,\tau\rangle \approx 
|J,K=J,M=J,\tau \rangle$ as the $a^{J,h_J,\tau}_J$ coefficient in ~\eqref{eq:asymwf2} dominates. After short algebra we find an 
explicit form of the asymmetric top wavefunction as
\begin{equation}
\begin{split}
\psi(t) = \frac{1}{4\pi}\sum_{J =J_{min}}^{J_{max}} \sqrt{2J+1}c_J e^{-i\phi J}\sum_{K=0}^J a^{J,h_J,\tau}_K \times\\
\times \left[e^{-iK\chi}d^J_{JK}(\theta) + (-1)^{\tau}e^{iK\chi}d^J_{J,-|K|}(\theta)\right]
\end{split}
\label{eq:explicitasymwf}
\end{equation}
where we note that the expansion coefficients $c_J$ are time-dependent and complex and can take the form: $c_J = 
|c_J|e^{-i(E_Jt+\phi^0_J)}$, where $\phi^0_J$ is the static phase of component state $|J_{K_{\lambda}\approx J}, 
M=J,h_J,\tau\rangle$ and $E_J$ is the energy of state. 

The alignment cosine measuring the degree of alignment with respect to the $X$ axis in the laboratory frame is given by
\begin{equation}
\begin{split}
\langle \cos^2 \phi \rangle_{\psi(t)} = \langle \psi(t) | \cos^2 \phi| \psi(t)\rangle
\end{split}
\label{eq:cosphi1}
\end{equation}
where $\phi \in [0,2\pi )$ is the polar Euler angle in the $XY$ plane. With the wavefunction given in~\eqref{eq:explicitasymwf} the 
alignment cosine from~\eqref{eq:cosphi1} can be further written as

\begin{widetext}
	\[
\begin{split}
\langle \cos^2 \phi \rangle_{\psi(t)}  &=  \frac{1}{16\pi^2} \sum_{J,J' =J_{min}}^{J_{max}} c_J^* c_{J'} \sqrt{(2J+1)(2J'+1)}\times 
\int_{0}^{2\pi} 
d\phi \cos^2\phi e^{-i\phi \Delta J} \sum_{K,K'=0}^{J,J'} a^{J,h_J,\tau*}_K a^{J',h_J',\tau'}_{K'}\times \\
 & \times \int d\Omega \left[
e^{-i\Delta_- K\chi}d^{J'}_{J'K'}d^{J}_{JK}+e^{i\Delta_- K\chi}d^{J'}_{J',-K'}d^{J}_{J,-K}+ 
(-1)^{\tau}(e^{i\Delta_+ K\chi}d^{J'}_{J',-K'}d^{J}_{J,K}
+e^{-i\Delta_+ K\chi}d^{J'}_{J',K'}d^{J}_{J,-K})
\right]
\end{split}
	\]
\end{widetext}
where $\Delta J = J'-J$, $\Delta_{\pm} K = K'\pm K$ and $d\Omega = \sin\theta d\theta d\chi$. Integrals over the $\phi$ angle 
impose 
rigorous selection rules on total angular momentum coupled by the $\cos ^2\phi$ operator:
\begin{equation}
\begin{split}
\int_{0}^{2\pi} 
d\phi \cos^2\phi e^{-i\phi \Delta J}  =   
\begin{cases}
2\frac{\sin(\pi \Delta J)}{\Delta J}\frac{\Delta J^2 -2}{\Delta J ^2 -4} e^{-i\pi\Delta J},& \text{if } \Delta J\neq  2\\
\frac{\pi}{2},              & \text{if} \; \Delta J = 	2
\end{cases}
\end{split}
\label{eq:intphi}
\end{equation}
we note that the integral given in~\eqref{eq:intphi} is non-zero only if $\Delta J = 2$. The integral over the $\chi$ Euler angle can 
also be carried out analytically:
\begin{equation}
\begin{split}
\int_{0}^{2\pi} 
d\phi  e^{-i\phi \Delta_{\pm} K}  =   
\begin{cases}
-\frac{i}{\Delta_{\pm}K}(1-e^{-i2\pi\Delta_{\pm}K}), & \text{if }\Delta_{\pm} K\neq  0\\
2\pi,              & \text{if} \; \Delta_{\pm} K = 0
\end{cases}
\end{split}
\label{eq:intchi}
\end{equation}
whereas appropriate integrals over the azimuthal Euler angle $\theta$ are generally denoted as:
\begin{equation}
b_{JJ'KK'} = \int_0^{\pi} d\theta \sin\theta d^{J}_{J,K}(\theta) d^{J'}_{J',K'}(\theta)
\label{eq:inttheta}
\end{equation}
Equation ~\eqref{eq:intchi} restricts contributions to the alignment cosine from states with the same $K$ quantum number. After 
some algebra one finds explicit expression for the alignment cosine $\langle \cos^2 \phi \rangle_{\psi(t)} $:
\begin{widetext}
	\[
	\begin{split}
	\langle \cos^2 \phi \rangle_{\psi(t)}  &=  \frac{1}{8} \sum_{J =J_{min}}^{J_{max}-2} |c_J| |c_{J'}| \sqrt{(2J+1)(2J+5)}\times 
\left[2 b_{JJ+200}(1+(-1)^{\tau})Re( a^{J,h_J,\tau*}_0 a^{J+2,h_{J+2},\tau}_0)+\right. \\
&\left.+ \sum_{K\neq 0}^{J} Re( a^{J,h_J,\tau*}_K 
a^{J+2,h_{J+2},\tau}_K)(b_{JJ+2KK} + b_{JJ+2-K-K})
	\right] \cos(\omega_{J+2J}t + \Delta \phi^0_{J+2J})
	\end{split}
	\]
\end{widetext}
If we assume that only states with $K=J$ contribute significantly to the wavepacket, we arrive at the equation 
~\eqref{eq:cogcosine} from the main text, where we note that $Re( a^{J,h_J,\tau*}_J
a^{J+2,h_{J+2},\tau}_{J+2}) \approx 1$,  $d^J_{JJ}(\theta) = \cos^{2J}\frac{\theta}{2}$, $d^J_{J,-J}(\theta) = 
\sin^{2J}\frac{\theta}{2}$ and $b_{JJ+2JJ} =  b_{JJ+2-J-J} = \frac{2}{2J+3}$. 

\bibliography{cmi,b-rotation}
\end{document}